# LensPlus: A High Space-bandwidth Optical Imaging Technique

Neha Goswami[1,a], Mark A. Anastasio[1,b]
Department of Bioengineering, University of Illinois Urbana-Champaign, Illinois, 61801, USA
[a]nehag4.illinois@gmail.com, [b]maa@illinois.edu

**Abstract**
The space-bandwidth product (SBP) imposes a fundamental limitation in achieving high-resolution and large field-of-view image acquisitions simultaneously. High-NA objectives provide fine structural detail at the cost of reduced spatial coverage and slower scanning as compared to a low-NA objective, while low-NA objectives offer wide fields of view but compromised resolution. Here, we introduce LensPlus, a deep learning-based framework that enhances the SBP of quantitative phase imaging (QPI) without requiring hardware modifications. By training on paired datasets acquired with low-NA and high-NA objectives, LensPlus learns to recover high-frequency features lost in low-NA measurements, effectively bridging the resolution gap while preserving the large field of view thereby increasing the SBP. We demonstrate that LensPlus can transform images acquired with a 10x/0.3 NA objective (40x/0.95 NA for another model) to a quality comparable to that obtained using a 40x/0.95 NA objective (100x/1.45NA for the second model), resulting in a 2D-SBP improvement of approximately 3.5x (2.04x for the second model). Importantly, unlike adversarial models, LensPlus employs non-generative model to minimize image hallucinations and ensure quantitative fidelity as verified through spectral analysis. Beyond QPI, LensPlus is broadly applicable to other lens-based imaging modalities, enabling wide-field, high-resolution imaging for time-lapse studies, large-area tissue mapping, and applications where high-NA oil objectives are impractical.

## 1. Introduction

Optical super resolution microscopy has seen a tremendous growth in fluorescence-based measurement techniques using both hardware developments [1-6] and recent machine learning based enhancements [7-10]. However, QPI [11-14], in general, still suffers from diffraction-limited transverse resolution. The source of this limitation is the imaging optics and the associated spatial frequency cutoffs [14]. Researchers have been pushing the limits of achievable resolution in QPI through both hardware innovations and computational strategies [15, 16]. In the hardware developments, structured illumination and synthetic apertures using oblique illuminations are among the principal approaches for resolution enhancement in QPI. These systems employ frequency modulation during acquisition and subsequent demodulation to reconstruct resolution-enhanced images [17-23]. In structured illumination, the spatially modulated illumination is used to enhance the frequency coverage of the imaging system by shifting the sample's high frequencies into the passband of the imaging system. This shift is proportional to the modulation frequency of structured illumination [19]. In synthetic aperture microscopy, the high-frequency information of the sample is made to fall into the existing aperture-based frequency range by using oblique illumination [24, 25]. Such oblique illuminations can be achieved using an array of illumination LEDs or digital micromirror devices. Diffraction-limited information from different angular illuminations is then combined into a single image with resolution superior to that of the individual components [24]. Techniques using speckled illumination are also reported to increase the resolution in phase imaging [26].

On the software end, deconvolution microscopies have demonstrated strong potential to enhance the resolution without hardware modifications [27]. The performance of such deconvolution techniques depends on the estimate of the instrument point spread functions (PSF) which can be theoretically modelled and fitted with experimental measurements. Deep-learning has also yielded encouraging results for resolution enhancement in phase imaging [28].

While resolution enhancement beyond the diffraction limit is one of the most pursued research topics in super resolution QPI, there exists another problem worth solving: increasing the space-bandwidth product (SBP) of the QPI imaging system [29, 30]. For a lens-based imaging system, field of view (space) and resolution (span of k-vectors, bandwidth) form a pair of negatively correlated degrees of freedom, as shown in Fig. 1. An increase in magnification brings an increase in resolution at the cost of a lower field of view and vice-versa. This is particularly problematic when using a high NA oil-based objective for which time-lapse imaging and large field of view scanning becomes difficult. SBP improvements using hybrid approaches (both hardware and software) have been reported in the literature [30]. Wang et. al [31] demonstrated an increased SBP for a coherent imaging system using a stack of defocused images along with computational reconstruction. This method requires additional optics: a demagnification camera adaptor and a stack of 3 to 5 defocused images per reconstruction.

In this paper, we introduce a deep learning-based method to increase the space-bandwidth product in spatial light interference microscopy (SLIM) [32] using a single quantitative phase image and without introducing additional hardware. We call this method LensPlus as it enables us to use a low-NA objective lens with an increased resolution comparable to a high NA objective, while preserving the wide field of view intrinsic to low NA objectives. Liu et al. [28], introduced similar work earlier that can be regarded as an early precursor to our approach. They used a generative-adversarial network (GAN) to translate an image acquired using a 4x/0.13NA objective lens to that equivalent to a 10x/0.3NA. The same group also demonstrated a similar method for fluorescence microscopy using GANs [33]. Another study [34] also employed GANs to enhance resolution of QPI images for higher resolution as compared to the previously mentioned study [28]. Our work expands significantly over these prior studies (Supplementary Table 1) and has several advantages in terms of result quantification, the enhancements to higher resolutions-for widely used objective lenses (40x/0.95NA and 100x/1.45NA) and the absence of a generative-adversarial model, thereby minimizing the risk of hallucinated features as evidenced by the spectral analysis of our results. We also adapt the commonly used image comparison metrics, namely peak signal to noise ratio (PSNR) and multiscale structural similarity index measure (MSSSIM), to incorporate biological content-awareness, leading to the introduction of two gradient sensitive metrics-gradient sensitive PSNR and gradient sensitive MSSSIM.

The significance of LensPlus lies in its ability to improve SBP without modifying the hardware while also providing a computationally rigorous performance evaluation of the models and eliminating the risk of hallucinations using a non-generative model. The method, although demonstrated for QPI images, can be translated to any other lens-based microscopy technique. It can also be used to eliminate the need for high-NA oil immersion objectives for time-lapse and wide-area scans at a comparable to high-NA resolution.

## 2. Methods

### 2.1 Sample preparation, image acquisition, registration, and preprocessing

The imaging modality used in this work is Spatial Light Interference Microscopy (SLIM). It is a quantitative phase microscopy based on interference of a temporally phase modulated reference beam and scattered beam from the object [32]. The controlled temporal phase modulation allows for the extraction of phase delay introduced by the sample and hence the compositional details (refractive index coupled with depth fluctuations in the sample) [14, 32]. It is an external add-on module (CellVista SLIM Pro, PhiOptics Inc.) attached to the output port of an inverted Nikon Eclipse Ti microscope. The objectives are phase contrast objectives (Ph1 10X, Ph2 40X and Ph3 100X ). The exposure time per frame was 10ms, 30ms and 50ms for 10x, 40x and 100x acquisitions. Each pixel in the output image represents this quantitative phase information expressed in radians. Following the pipeline as shown in Fig. 1, pairs of low-resolution (input) and high-resolution (target) images were acquired as detailed below.

The samples for this work were fixed HeLa cells plated on 6-well glass bottom plates. HeLa cells (ATCC CCL-2) were thawed from cryopreserved vials and expanded in T-75 flasks using DMEM supplemented with 10% FBS and 1% penicillin–streptomycin at 37 °C and 5% $CO_2$. After being subcultured three times to ensure healthy growth, the cells were transferred to 6-well glass-bottom plates for experimental preparation. Once they reached the appropriate confluence after 48 hours of growth, cells were fixed in 4% paraformaldehyde for 30 min at room temperature, washed three times with PBS, and stored in PBS at 4 °C until imaging. SLIM imaging was performed (additional details in the supplementary text) with 10x, 40x and 100x for the training data. For 10x to 40x (10xplus model) training, training patches of 384 x 384 pixels were extracted from the 10x image and resized to 1536 x 1536 pixels using bilinear interpolation to match the 1536 x 1536 pixels high-resolution 40x image. Phase correlation-based registration was then performed in MATLAB to register the images. Background subtraction was performed on the target high-resolution images. 512 x 512 cropped image patches were then extracted from both the input and target images for training. The size of the training, validation and test dataset was then: 10817, 1923 and 2500 images respectively.

For 40x to 100x (40xplus model) training, patches of 850 x 850 were extracted from 40x image and upscaled to 2125 x 2125 image using bilinear interpolation. 40x and 100x images were then registered, background-subtracted, and cropped into 512 x 512 patches for training, yielding 10052, 3933 and 4084 images for training, validation and test respectively.

For the 10xplus model, for the efficient use of available dataset and for data augmentation, an additional step was involved: the training patches contained partial overlap at the edges, but the validation dataset did not have such edge overlap-the non-overlapping crops were extracted from the image and the edges where the pixels were less than 512, an empty phase strip was included in such edges from the empty patches. The test set also contained the edge overlaps, but the pixel-metric was not evaluated on the edges, only the central non-overlapping area per patch (removing 100 pixels per edge) was used for these evaluations. The data for 40xplus model was acquired without such overlaps, however the test evaluations were again performed on the central regions (removing 50 pixels per edge) to maintain consistency in evaluation. For spectral analysis, however, we did not remove the border pixels. Input images were normalized using global minima and maxima values of -2 and 2 for 10xplus and -1.5 to 1.5 for 40xplus. These bounds were empirically determined to match the typical dynamic range of SLIM phase images at each magnification for the monolayer, single-cell samples. Preprocessing was performed in ImageJ (FIJI) and MATLAB.

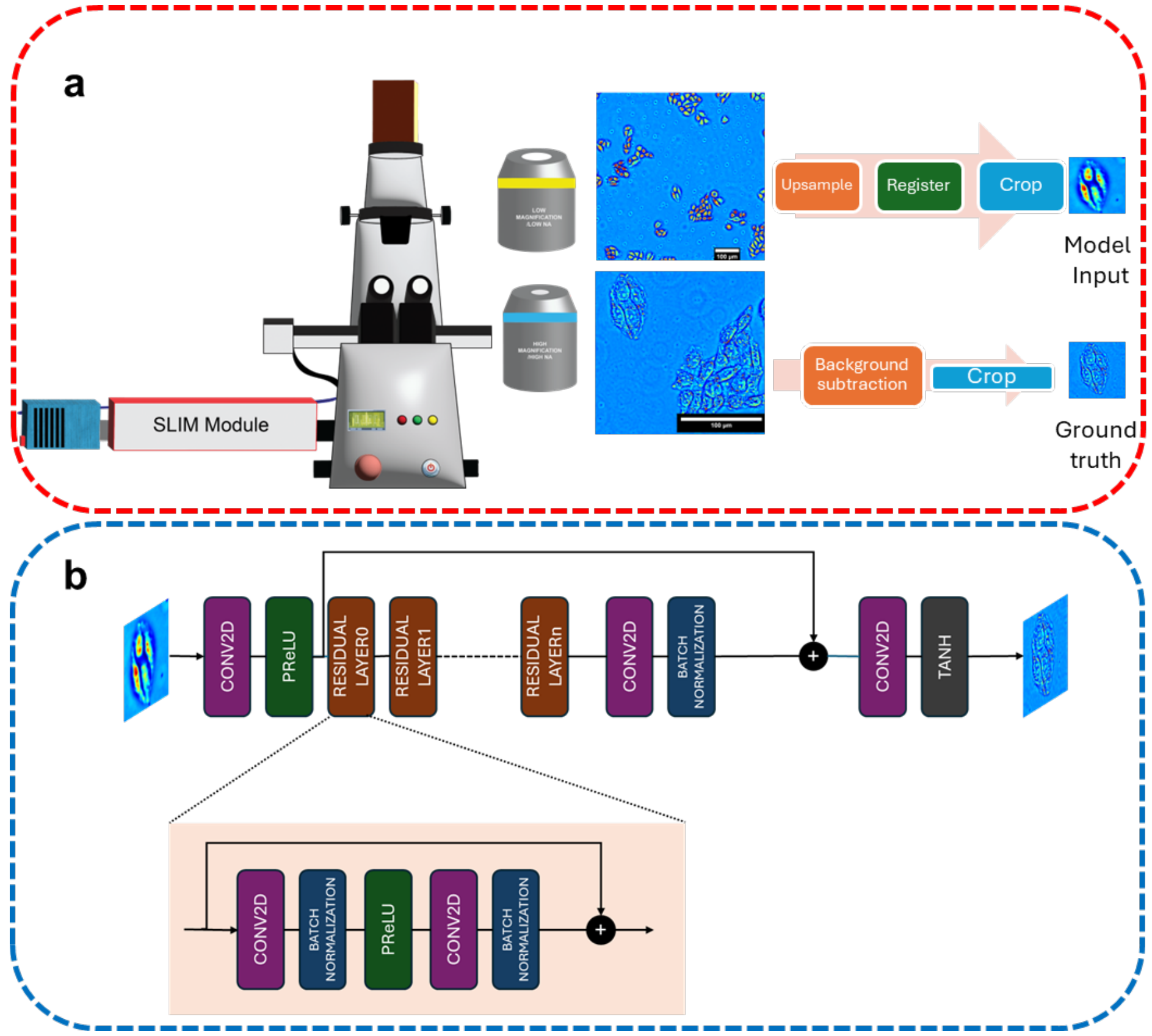

Figure 1. Workflow a. Schematic of the microscopy system with SLIM module attached to the output port. The corresponding representative fields of view imaged using a low magnification/low NA objective lens (top) and a high magnification/high NA objective lens (bottom). Low resolution images (top) are upsampled, registered and cropped to prepare input patches for training, while high resolution images (bottom) are processed to remove background and then cropped to prepare ground truth target patches for model training. b. Modified super resolution residual network (SRResNet) architecture with n+1 (8 for 10xplus and 16 for 40xplus model) residual layers (expanded in inset).

### 2.2 Modified SRResNet models and training details

The 10xplus and the 40xplus models are based on modified versions of SRResNet [35]. One key distinction from the original SRResNet architecture is the removal of pixel-shuffling layers. This decision was motivated by the fact that input images were already bilinearly upsampled to match the resolution of the target images, making pixel-shuffling unnecessary. We used 8 and 16 residual units for the 10xplus and 40xplus models respectively. The final activation was the *tanh* function. The ground truth images were correspondingly normalized to the [-1, 1] range. The model is trained with a combination loss comprising MSE loss and Pearson correlation coefficient loss, defined as

$$L_{MSE} = E\left[\left(y-\hat{y}\right)^2\right] \quad . \tag{1}$$

The Pearson loss (PCL) is defined as:

$$L_{PC} = E\left[\left(1-\frac{\sum\left(y-\bar{y}\right)\left(\hat{y}-\bar{\hat{y}}\right)}{\sqrt{\sum\left(y-\bar{y}\right)^2\sum\left(\hat{y}-\bar{\hat{y}}\right)^2}}\right)\right] \quad . \tag{2}$$

The combined loss is:

$$L = \varepsilon_1 L_{MSE} + \varepsilon_2 L_{PC} \quad . \tag{3}$$

With $\varepsilon_1 = 1$ and $\varepsilon_2 = 0.4$ for MSE and PCL respectively for 10xplus and $\varepsilon_1 = 1$ and $\varepsilon_2 = 0.6$ for 40xplus models. The Adam optimizer was used for training, with $\beta_1 = 0.8$ and $\beta_2 = 0.7$ is used for model optimization. Training was terminated after no-major changes in the validation loss were observed and the model with lowest validation loss (at epochs=74 for 10xplus and 96 for 40xplus) was selected as the final model, results shown in Figs. 3 and 4 for the 10xplus and 40xplus models respectively. Batch size was set to 2. The training strategy and associated parameters remain the same except for the learning rate, which was 5e-6 for 10xplus and 1e-6 for 40xplus model. The model architecture is shown in Fig. 1b, with the inset showing the constituents of the residual block. The model was developed upon the base SRResNet architecture as provided in Ref. [36]. All the relevant codes were developed using PyTorch and Python. Training till termination on an NVIDIA RTX A6000 GPU for 10xplus model (dataset with 10817, 1923 and 2500 training, validation and test images of size 512 x 512) took 30 hours and for the 40xplus model (dataset with 10052, 3933 and 4084 images of size 512 x 512) took 52.56 hours.

### 2.3 Image content estimation and content-aware pixel-metrics

The image content was quantified by summing the gradient magnitudes in both x and y directions per test image patch. The gradient magnitude was computed using MATLAB's in-built *imgradient* function. The summed gradient magnitude was then normalized over the whole test dataset to obtain the final measure. Mathematically, this operation can be described as:

$$G_n = \sum_{xy}\left(\sqrt{\left(\frac{\partial I_n}{\partial x}\right)^2+\left(\frac{\partial I_n}{\partial y}\right)^2}\right), \tag{4}$$

Where $G_n$ is the gradient sum for $n^{th}$ image $I_n$, x and y are the orthogonal dimensions of the image. The normalized measure for the entire test dataset is then:

$$g_n = \frac{G_n}{\max_n(G_n)}. \tag{5}$$

The representative input images for different values of $g_n$ for both 10xplus and 40xplus models are shown in Fig. 2a and b respectively. It can readily be seen that as this measure approaches 1, the true content in the image increases, while lower values are indicative of blank or background-dominated regions

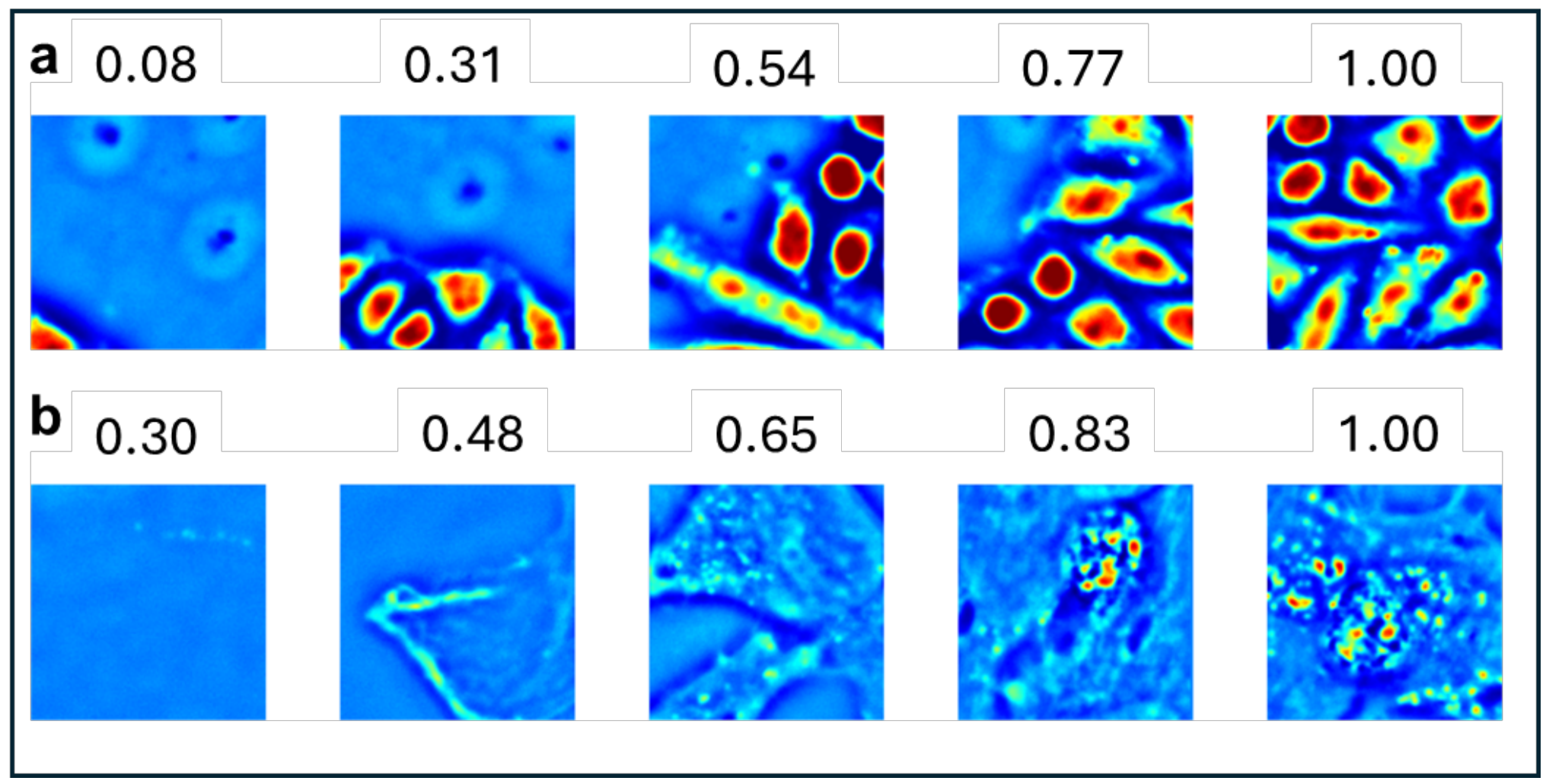

Figure 2. Content-estimation: representative images for increasing normalized gradient sum values for a. 10x/0.3NA images b. 40x/0.95 NA images.

The content aware pixel metrics are then formulated as the weighted versions of the original metric definitions, the weighing factor being $g_n$.

The traditional definition of PSNR is

$$PSNR = 10\log_{10}\left(\frac{p^2}{mse\left(y,\hat{y}\right)}\right), \tag{6}$$

Where, $p$ is the peak value in the image based on the data-type and $mse$ denotes the mean squared error between ground truth $y$ and prediction $\hat{y}$.

The gradient-sensitive PSNR or GraS-PSNR is then defined as:

$$GraS-PSNR = PSNR * g_n. \tag{7}$$

Similarly, the MSSSIM is defined as [37]

$$MSSSIM = \left[l_M\left(y,\hat{y}\right)\right]^{\alpha_M} . \prod_{j=1}^{M}\left[c_j\left(y,\hat{y}\right)\right]^{\beta_j}\left[s_j\left(y,\hat{y}\right)\right]^{\gamma_j}, \tag{8}$$

where the luminance, contrast, and structure components are denoted by *l, c*, and *s,* as defined in Ref. [37]

The gradient-sensitive MSSSIM or GraS-MSSSIM is then defined as:

$$GraS-MSSSIM = MSSSIM * g_n. \tag{9}$$

For the spectral analysis, the spatial Fourier transform of each image was calculated, and from this the power spectral density was extracted and radially averaged over binned frequency ranges for each model. The mean and standard deviation was calculated across the test images, and the resultant curve was normalized by the maximum mean value. The 3dB bandwidth was calculated for the resultant mean curves as the frequency range over which the normalized radially averaged power spectral density falls to half of its peak value. The gain was calculated as:

$$G_{pw}(\%) = 100\frac{\Delta k_{pred} - \Delta k_{input}}{\Delta k_{gt} - \Delta k_{input}} \tag{10}$$

We also employed established evaluation technique from super-resolution microscopy: Fourier Ring Correlation (FRC) [38, 39]. For the FRC calculation, the following expression was implemented [38, 39]

$$FRC(k_i) = \frac{\sum_{k\in k_i} \widetilde{I}_1(k).\widetilde{I}_2^*(k)}{\sqrt{\sum_{k\in k_i}\left|\widetilde{I}_1(k)\right|^2 . \sum_{k\in k_i}\left|\widetilde{I}_2(k)\right|^2}}. \tag{11}$$

where, $k_i$ is the $i^{th}$ frequency ring, $\tilde{I}$ is the Fourier transform of the image and 1, 2 represents two different resolution images of the same object.

FRC curves for the test set were averaged resulting in Figs. 5g and 6g for 10xplus and 40xplus models respectively, where the central solid lines indicate the mean and shaded portions denote the standard deviation. The effective resolution/spectral bandwidth was estimated by following the 1/7 criterion [38, 39]. The gain was defined as:

$$G_{frc_bw}(\%) = 100\frac{\Delta k_{gt:pred} - \Delta k_{gt:input}}{\Delta k_{gt:input}} \tag{12}$$

For the other metrics, the gain was defined as % change between before and after values in each of the tables (Tables1 and 2). For calculation of all the metrics, images with low content/background were removed prior to the metric calculations as detailed in the Results section. Model performance quantification analysis, post-prediction, was performed in MATLAB.

## 3. Results

### 3.1 Enhanced image resolution for two pairs of objectives

Modified versions of the standard SRResNet [35] model were developed and employed for two distinct aims: translation from 10x/0.3NA to 40x/0.95NA (henceforth referred to as the 10xplus model) and translation from 40x/0.95NA to 100x/1.45NA (henceforth referred to as the 40xplus model). The results of the 10xplus model (see Methods Section for model details), are shown in Fig. 3. Full field of view acquired using a 10x/0.3NA objective is shown in Fig. 3a, with the red rectangular area (enlarged in Fig. 3b) illustrating the corresponding full field of view of a 40x/0.95NA objective lens, shown in Fig. 3c. Representative cropped patches from the hold-out test set are shown in Fig. 3d (10x-input), with the corresponding ground truth patches in Fig. 3e. The model predictions are shown in Fig. 3f, where the enhancements of resolution are clearly visible. The colorbar indicates phase delay in radians.

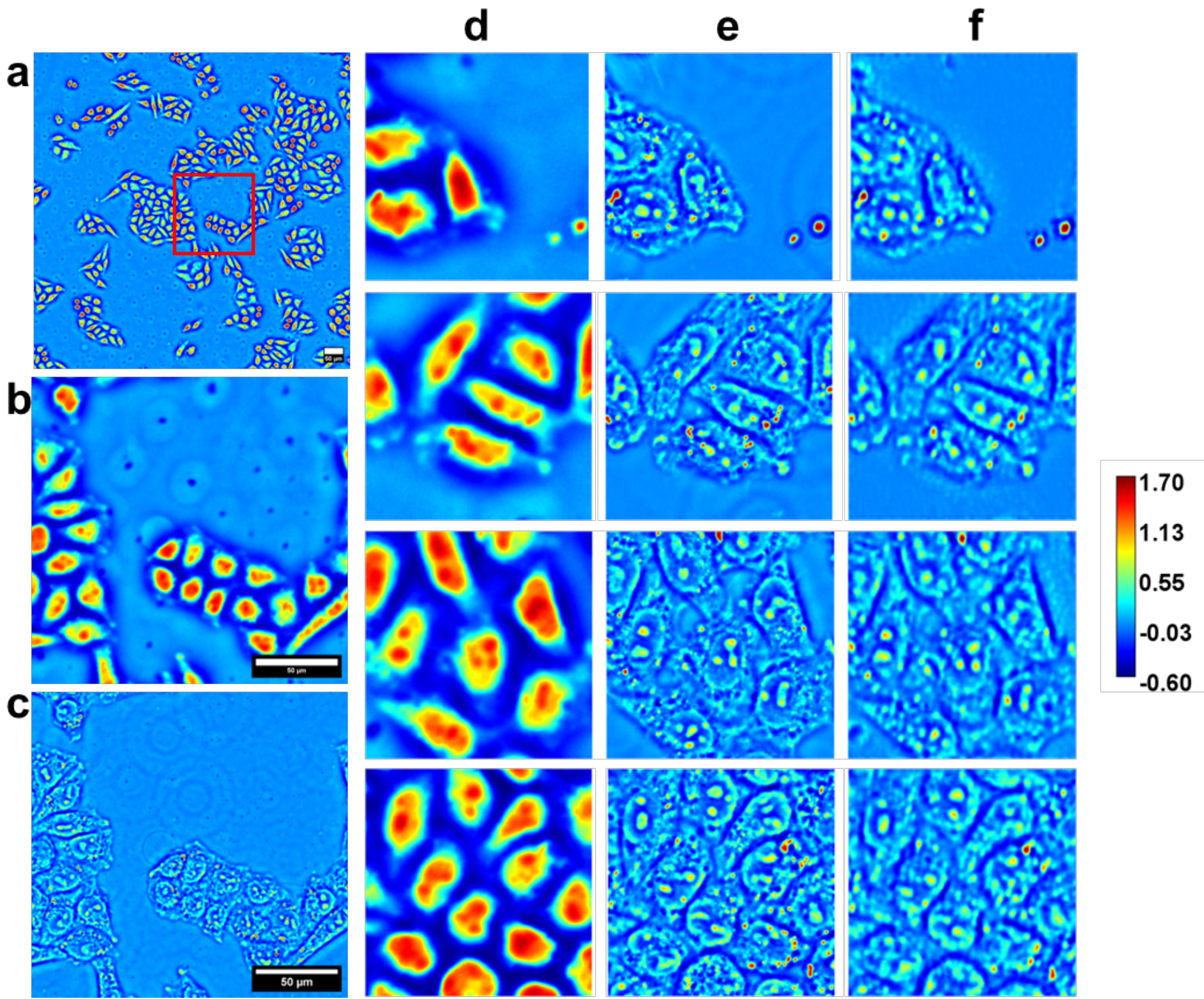

Figure 3. 10xplus model performance. Representative images for a. full field of view of a low-resolution (10x/0.3NA) objective, b. magnified region corresponding to the red rectangle in a, c. corresponding high-resolution image (40x/0.95NA). d. test, low-resolution (10x/0.3NA) input images, e. corresponding high-resolution (40x/0.95NA) ground truth images , f. model predictions showing resolution enhancement. The colorbar indicates phase delay in radians. Scalebar measures 50 µm in a-c.

The 40xplus model was trained for even higher targeted resolution enhancement, from 40x/0.95NA to 100x/1.45NA. Representative images for the full field of view for each objective are shown in Fig. 4a for 40x, 4c for 100x and the cropped region inside the red rectangle in 4a is shown in 4b. The model performance on a hold-out test dataset (Fig. 4d input, 4e ground truth) is shown in Fig. 4f.

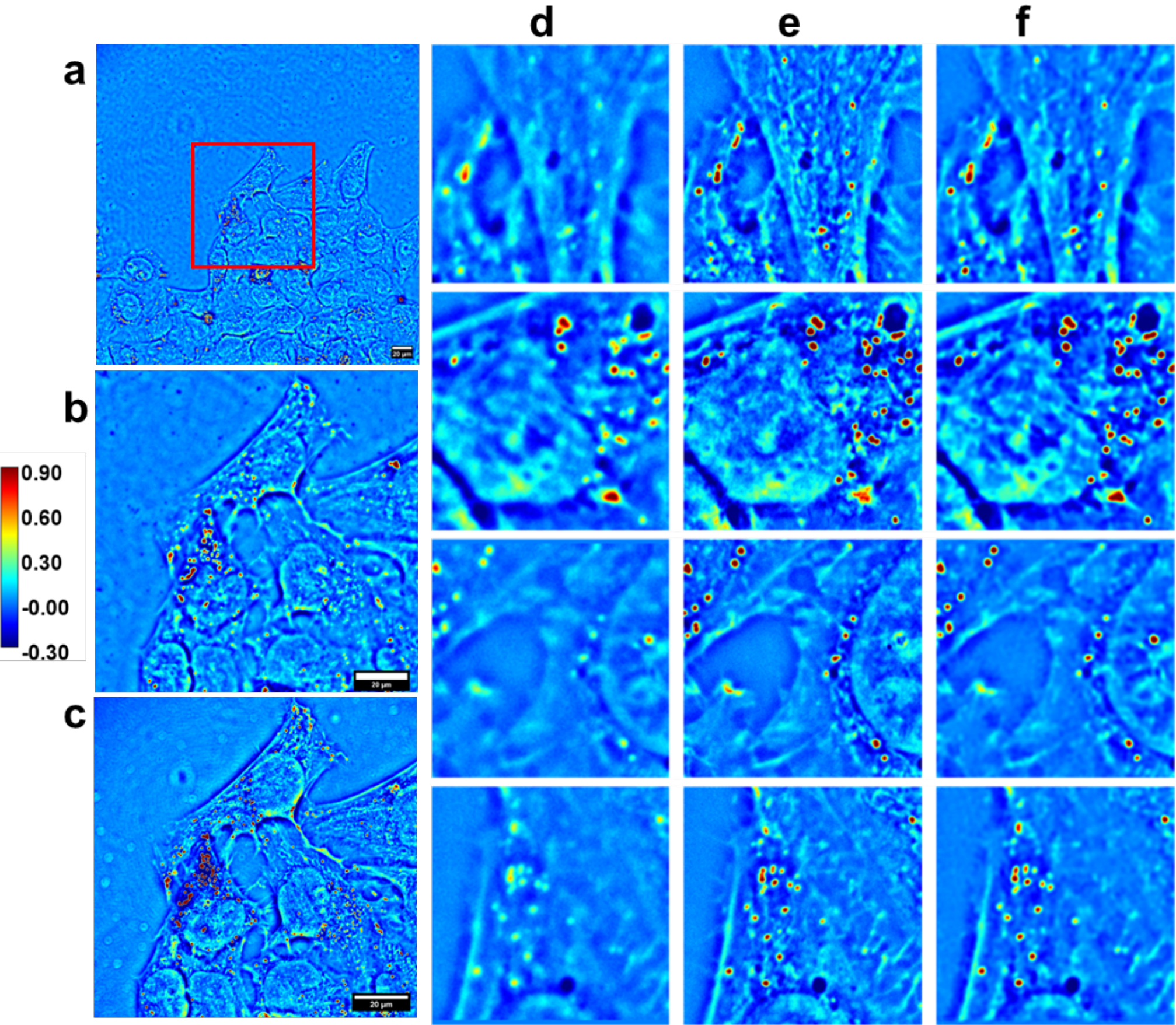

Figure 4. 40xplus model performance. Representative images for a. full field of view of a low resolution (40x/0.95NA) objective, b. magnified region from the red rectangle in a, c. corresponding high resolution image (100x/1.45NA). d. test, low-resolution (40x/0.95NA) input images, e. corresponding high-resolution (100x/1.45NA) ground truth images , f. model predictions demonstrating enhanced resolution features. Colorbar indicates phase delay in radians. Scalebar measures 20 µm in a-c.

### 3.2 Quantitative evaluation of LensPlus

Both the models (10xplus and 40xplus) were initially evaluated over a hold-out dataset using traditional image quality metrics such as Peak Signal-to-Noise Ratio (PSNR), Multiscale Structural Similarity Index Measure (MSSSIM) [37] . The results are shown in Fig. 5 (for 10xplus) and Fig. 6 (for 40xplus) models as a function of image content quantified using the normalized image gradient sum (see Methods Section). Interestingly, we found an inverse relationship between both PSNR and MSSSIM and the image content as illustrated in Figs. 5a,b and Figs.6a,b, with higher content images showing lower metrics and lower content (mostly blank/background) images showing higher metrics, highlighting the limitations of these metrics for evaluating resolution enhancement of biological images arising from different sources.

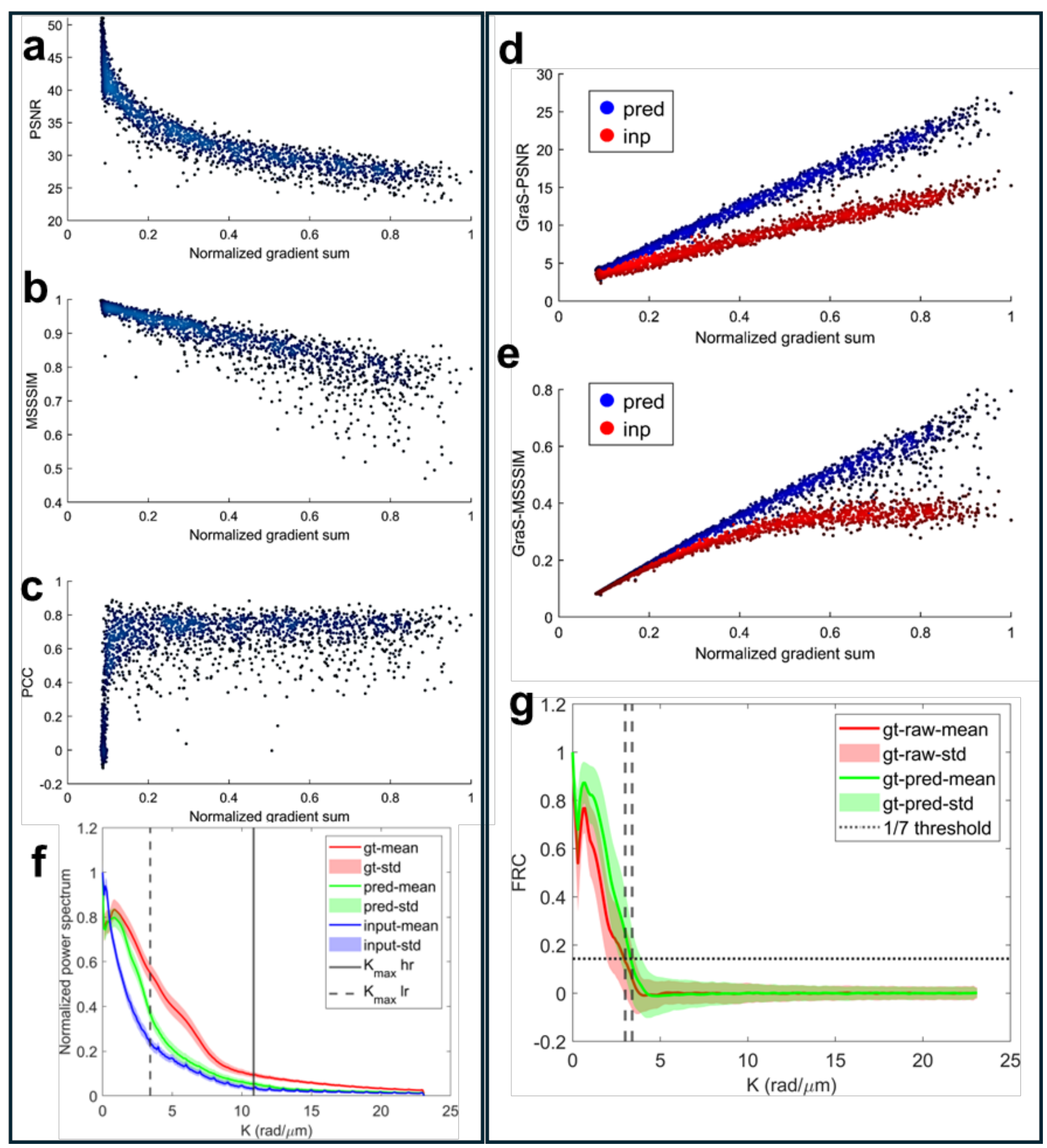

Figure 5. Quantitative Performance of 10xplus. a. PSNR b. MSSSIM c. PCC vs normalized gradient sum for test dataset. Content-aware metrics: d. GraS-PSNR, e. GraS-MSSSIM for the same test dataset with blue curve showing the prediction vs gt and red curve showing input vs gt trends. Spectral metrics: f. normalized radially averaged power spectral density averaged over the test dataset with solid curves denoting the mean and shaded bands representing standard deviations, blue curve: raw (input), red curve: ground truth target, green curve: prediction, the first vertical dotted line is the theoretical frequency cut-off for 10x/0.3NA objective and the second vertical dotted line is the theoretical frequency cut-off for 40x/0.95NA objective. g. Fourier ring correlation (FRC) curve averaged over test dataset, with solid curves denoting the mean and shaded bands representing standard deviations, red curve represents raw (input) vs ground truth FRC and green curve represents prediction vs ground truth FRC, horizontal dotted line represents the 1/7 resolution threshold, with vertical dotted lines representing the corresponding frequency cut-offs.

The Pearson correlation coefficient (PCC) on the other hand showed a relatively flat response to the image content (Figs. 5c and 6c for 10xplus and 40xplus models respectively), demonstrating its utility as a metric for this problem. The initial bar in the PCC curve in both 10xplus (Fig. 5c) and 40xplus (Fig. 6c) provided a good normalized gradient sum threshold for removing the almost blank/empty images from the test set such that the pixel metrics are not inflated. Using these cutoff (0.15 for 10xplus and 0.4 for 40xplus), we refined our test set. The mean PSNR, MSSSIM and PCC for these information rich frames are now: 30.57, 0.86 and 0.69 for the 10xplus model (Table 1) and 34.18,0.92 and 0.83 for the 40xplus model (Table 2). Compared to the input, the gain in these metrics is 49.85%, 26.53% and 190.12% for the 10xplus model (Table 1) and 4.24%, 2.57% and 5.89% for the 40xplus model (Table 2).

### 3.3 Content-aware evaluation metrics: Gradient sensitive PSNR (GraS-PSNR) and MSSSIM (GraS-MSSSIM)

Although the obtained PSNR and MSSSIM metrics indicate success of our models' performance in the traditional computer vision paradigm, they still do not justify the inverse correlation with image content in our application where we compare images that have varying content across the test dataset. Further the two images (paired low resolution and high resolution) are not simply a degraded version of another-meaning low resolution images do not simply result from a noisy and blurred high resolution image-they represent two different measurements, one from each objective lens. This can introduce additional pixel-level differences owing to differences in aberrations in each lens and different sectioning capabilities. Here, we introduce content-aware image metrics which are a simple modification to the existing definition of the current metrics (see more in Methods Section). Gradient sensitive PSNR: GraS-PSNR and gradient sensitive MSSSIM: GraS-MSSSIM are shown in Figs. 5d,e and Figs. 6d,e for the 10xplus and the 40xplus models, respectively. The red curve indicates the metric evaluation on the input-target pair, while the blue curve shows evaluations on the prediction-target pair. These curves weigh the original metric by the image content and hence portray a more meaningful trend. We also realize that using the gain in a measure as the evaluation metric rather than a single number evaluated on the predictions is a more informative way of evaluating a model. The gain in these metrics (GraS-PSNR and GraS-MSSSIM) is now: 54.35% and 35.71% for the 10xplus model (Table 1) and 4.26% and 2.69% for the 40xplus model (Table 2).

| metric | Before (input vs ground truth) | After (prediction vs ground truth) | Gain % | Interpretation |
|---|---|---|---|---|
| PSNR | 20.4025 | 30.5742 | 49.85 | Pixel metric |
| MSSSIM | 0.6763 | 0.8557 | 26.53 | Pixel metric |
| PCC | 0.2388 | 0.6928 | 190.12 | Global Pixel metric |
| GraS-PSNR | 9.1879 | 14.1812 | 54.35 | Gradient weighed (content-aware) metric |
| GraS-MSSSIM | 0.2940 | 0.3990 | 35.71 | Gradient weighed (content-aware) metric |
| Spatial bandwidth difference (rad/µm) | 2.4 (=3.9-1.5) | 1.1(=3.9-2.8) | 54.17 | Proximity to 3dB spectral bandwidth of ground truth |
| FRC bandwidth (rad/µm) | 3 | 3.4 | 13.33 | Similarity to the ground truth in spectral space |

Table 1. Quantitative performance metrics for 10xplus model.

### 3.4 Spectral evaluations

These gains point to an important difference in the performance of the two models. The 10xplus model shows much higher gains than the 40xplus model. For the 10xplus model, the features to recover were not as fine as for the 40xplus model, where the input was already a high-resolution image and even higher resolution was targeted for recovery.

Hence, even though the percentage gain is much lower in these pixel metrics for the 40xplus model, the visual inspection in Fig. 4 reveals an enhancement in resolution and thus can point to the fact that the pixel-level metrics may underestimate the 40xplus model performance. This prompted us to look for other metrics that can justify the initial resolution conditioning of the input on the model evaluation.

Radially averaged power spectral density (PSD) curves in Figs. 5f and 6f, show that the prediction (green curve) is closer to the ground-truth high resolution (red curve) than the low-resolution input (blue curve). The curves are the normalized average of the individual image's radially averaged power spectral densities, with the shaded portion showing the normalized standard deviation. The dotted lines represent the theoretical cutoff frequency of each optical system and the curve beyond that dotted line represents mostly high frequency noise. As can be seen from Fig. 5f, the 10xplus model is able to recover content at higher frequencies in the physically allowed frequency space for that objective. The trailing edge of the green curve shows that the model is not hallucinating, otherwise it would show greater content than ground truth after the frequency cutoff (Kmax=3.43, 10.85 and 16.56 rad/µm for 10x/0.3NA, 40x/0.95NA and 100x/1.45NA objectives respectively).

Similarly, for the 40xplus model, as shown in Fig. 6f, the green curve (prediction) lies between the blue curve (low-resolution input) and the red curve (high-resolution target) within the valid frequency range for the 40x/0.95NA objective. Notice that the green curve (prediction) dips below the blue curve (input) in the frequency ranges after the theoretical maximum frequency cut-off, indicating a decrease in high-frequency noise (denoising). This is an additional effect that we introduced in the high-resolution 40xplus model, which was trained on background-removed input–target pairs as compared to 10xplus model where background was removed only for the target images but not the input. Again, the absence of hallucinations is indicated by the flat green (prediction) curve after the frequency cut-offs.

We quantified power spectral differences using the gain in proximity to the 3dB bandwidth of the target spectrum. Aligning with our visual perception of the model performances, this metric, an evaluation of the gain in the proximity to the 3dB spectral bandwidth of target high-resolution ground truth data is 54.17% for 10xplus model (71.8% of the ground truth bandwidth recovered) (Table 1) and 75% for 40xplus model (90.91% of the ground truth bandwidth recovered) (Table 2).

We also evaluated our models using Fourier Ring Correlation [38, 39] (see more in Methods Section), which essentially denotes the similarity of the image under test to the target high-resolution ground truth image in the discretized frequency bins-an established technique to estimate useful resolution. The results are shown in Figs. 5g and 6g for 10xplus and 40xplus models, respectively. The green curve denotes the normalized average FRC curve for prediction vs ground truth and the red curve denotes the same for input vs ground truth. The shaded portions denote the normalized standard deviations. The horizontal dotted curve is the 1/7 cutoff resolution thresholds that are used to characterize the meaningful frequency bandwidths for comparison [38, 39]. The metric in this case is the gain in the spectral bandwidth of the green curve vs the red curve, with the values being 13.33% for the10xplus model (Table 1) and 39.47% for the 40xplus model (Table 2).

Interestingly, spectral domain analysis provides insights that align with the visual assessments of model performance.

**3.5 Increased space bandwidth product**

The motivation behind this work was to retain the high scan area of the low-resolution objective lens (space) while increasing its resolution (bandwidth). When the 10xplus and 40xplus models are applied to the patches of the low-resolution input and the results are stitched back, the stitched output exhibits both wide field of view and improved resolution (Fig. 7). Figs. 7 a,e show one field of view acquired by the low-resolution objective for the 10xplus and 40xplus models respectively. The images were preprocessed as per the model requirements (more in Methods Section) and the predictions were stitched together as shown in Figs. 7b,f for 10xplus and 40xplus models respectively. Regions enclosed in red rectangles in Figs. 7b,f are shown in Figs. 7d,h (model predictions) to be compared with one single field of view of the high-resolution objectives in Figs. 7c,g ( 40x/0.95 NA and 100x/1.45NA respectively). An increase in space-bandwidth product is clearly observable in both model outputs. Since, for our input images, we are preserving the space (field of view), the SBP enhancement along each transverse spatial dimension is effectively represented by increased spectral bandwidth (Tables 1 and 2), corresponding to 1.87x (10xplus) and 1.43x (40xplus) gains leading to 2D SBP gains of 3.5x and 2.04x (for 10xplus and 40xplus respectively). These numbers, combined with previously discussed metrics point to two important functional regimes for our models: when the input resolution is low (as for the 10xplus model), the gain in bandwidth ratio and hence SBP is more, although the fraction of recovered ground-truth bandwidth is lower due to the greater initial disparity (and hence the proximity to the ground truth bandwidth is low). For the high-resolution input (as for 40xplus, where the model is translating from high to even higher resolution), the gain in bandwidth ratio and SBP is less. Although the absolute gain is smaller, the recovered portion of high-frequency content is greater due to the smaller initial gap.

| **metric** | **Before (raw image vs gt)** | **After (pred vs gt)** | **Gain %** | **Interpretation** |
|---|---|---|---|---|
| PSNR | 32.7920 | 34.1812 | 4.24 | Pixel metric |
| MSSSIM | 0.9013 | 0.9245 | 2.57 | Pixel metric |
| PCC | 0.7871 | 0.8335 | 5.89 | Global Pixel metric |
| GraS-PSNR | 16.7122 | 17.424 | 4.26 | Gradient weighed (content-aware) metric |
| GraS-MSSSIM | 0.4617 | 0.4741 | 2.69 | Gradient weighed (content-aware) metric |
| Spatial bandwidth difference (rad/µm) | 2.52 (=6.93-4.41) | 0.63 (=6.93-6.3) | 75.00 | Proximity to 3dB spectral bandwidth of ground truth |
| FRC bandwidth(rad/µm) | 7.98 | 11.13 | 39.47 | Similarity to the ground truth in spectral space |

Table 2. Quantitative performance metrics for 40xplus model.

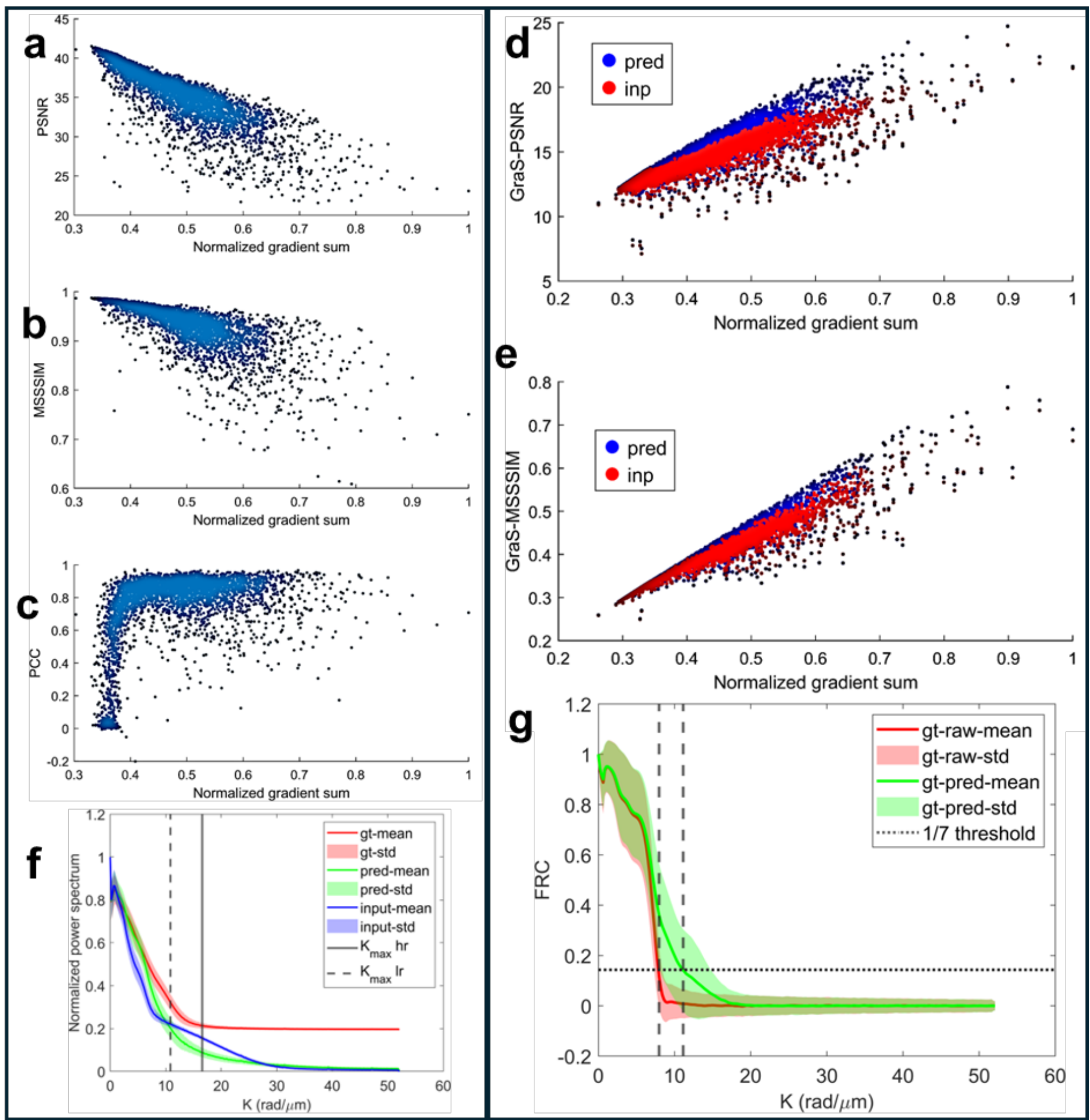

Figure 6. Quantitative Performance of 40xplus. a. PSNR b. MSSSIM c. PCC vs normalized gradient sum for test dataset. Content-aware metrics: d. GraS-PSNR, e. GraS-MSSSIM for the same test dataset with blue curve showing the prediction vs gt and red curve showing input vs gt trends. Spectral metrics: f. normalized radially averaged power spectral density averaged over the test dataset with solid curves denoting the mean and shaded bands representing standard deviations, blue curve: raw (input), red curve: ground truth target, green curve: prediction, the first vertical dotted line is the theoretical frequency cut-off for 40x/0.95NA objective and the second vertical dotted line is the theoretical frequency cut-off for 100x/1.45NA objective. g. Fourier ring correlation (FRC) curve averaged over test dataset, with solid curves denoting the mean and shaded bands representing standard deviations, red curve represents raw (input) vs ground truth FRC and green curve represents prediction vs ground truth FRC, horizontal dotted line represents the 1/7 resolution threshold, with vertical dotted lines representing the corresponding frequency cut-offs.

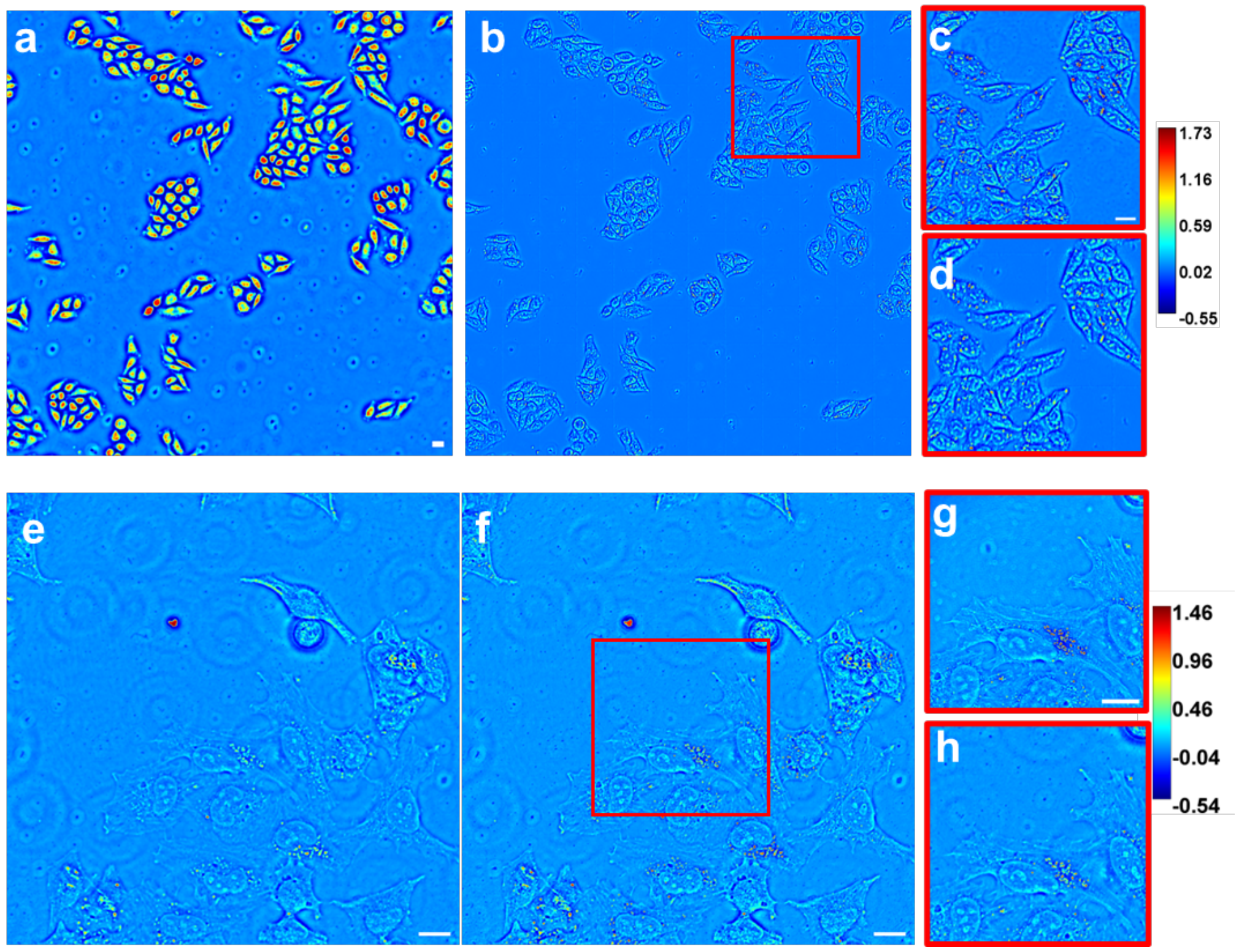

Figure 7. Increase in space-bandwidth product for (top) 10xplus and (bottom) 40xplus models. a,e. Full field of view for low NA objectives (10x/0.3NA and 40x/0.95NA respectively), b,f. corresponding model predictions with enhanced resolution. c,g. corresponding full-field of view ground truth, high-resolution target images (40x/0.95NA and 100x/1.45NA respectively) for the regions enclosed in red rectangles in panels b and f respectively, d,h. Zoomed in corresponding areas from b,f respectively showing resolution improvements. Scalebars 20µm. Color bars denote phase delay (in radians).

## 4. Discussion

In this work, we introduced LensPlus, a deep learning–based method for enhancing the space-bandwidth product (SBP) of lens-based optical imaging systems. While our implementation focused on spatial light interference microscopy (SLIM), the underlying approach is broadly applicable across other imaging modalities, including fluorescence and brightfield microscopy. We have demonstrated improved resolution and as a result an increase in space-bandwidth product for two magnifications 10x and 40x. Although, due to the physical limitations of the amount of information collected by low NA objectives, we could not possibly reconstruct all the frequencies that were missed by low NA objectives. However, we did gain an increase in resolution as evident in our results. We used a SRResNet model as the base model but made significant changes to it in terms of architecture, training loss and training strategy. A similar study was conducted by Liu et. al. [28], where they demonstrated a translation from 4x/0.13 NA to 10x/0.3NA using GAN. However, our translations from 10x/0.3NA to 40x/0.95 NA and 40x/0.95NA to 100x/1.45NA has a greater practical impact because of the frequency of usage of these objectives as compared to a 4x objective. Also, we did not employ GAN as used in other prior studies [34] because of the inherent issues of hallucinations [40, 41]. A comparison of our work with related work is shown in Supplementary Table 1. Other supervised deep learning models could be, in-principle, used in our framework, however we chose this model because it is well understood. Our key contribution is to show that SBP can be reliably increased, even when the problem is slightly ill-posed due to the sectioning differences between objectives, using the proposed method and our rigorous evaluation of the model performance and introduction of content-aware metrics suited for studies like ours. Future works can repeat this using more modern deep-learning approaches.

Conventional analytical SBP expressions in two-dimensional imaging scale with $NA^2$, reflecting the area of the passband in the k-space. The corresponding 2D SBP gain can be obtained by squaring our reported bandwidth ratios resulting in 3.5x and 2.04x for 10xplus and 40xplus models respectively.

A common assumption in many deep-learning-based super-resolution methods is that low- and high-resolution images represent identical spatial sections of a sample, differing only in resolution. However, in real, lens-based optical systems, especially in biological imaging, low-NA and high-NA objectives introduce fundamentally different axial point spread functions. High-NA lenses optically section more tightly, whereas low-NA lenses integrate signal over a broader axial depth. This means that paired images acquired from different objectives are not strictly identical in content, even when aligned in lateral space. The sectioning differences in the training pair can lead to differences in the predictions and the ground truth images. However, these differences are not artefacts or hallucinations because they are physically present in the input images due to their thicker optical sections. These effects are more prominent in the case of 10x/0.3NA to 40x/0.95NA translation as the difference in optical section is higher in this data than 40x/0.95NA to 100x/1.45NA. Despite this inherent mismatch, which results in an ill-posed problem, LensPlus can reconstruct higher-resolution images that are physically faithful to high-NA outputs, as evidenced by power spectral density (PSD) and Fourier ring correlation (FRC) analyses. This is because the involved deep learning model is able to restrict the solution space via the inductive bias of the network and implicitly learning priors that describe the high-resolution images. These results suggest that our model has learned to account for axial integration effects in a data-driven manner, without the need for explicit 3D modeling.

We acknowledge that our training strategy uses the best-focused 2D slice from each objective. While this does not fully resolve the sectioning disparity, it reflects the standard practice in microscopy, where users typically acquire a single focused image rather than full 3D stacks. This design choice allows us to frame LensPlus as a 2D-to-2D enhanced-resolution tool, enabling deployment in high-throughput, wide-area scanning applications without increasing acquisition burden. Methods using z-stack projections or full volumetric imaging may better model axial mismatch but depart from the practical 2D imaging scenario we aim to improve.

One key strength of LensPlus is its purely data-driven design. In contrast to physics-based or hybrid methods [42] that rely on convolutional kernel assumptions (often Gaussian), our model makes no a priori assumptions about the system's point spread function. While physics-informed methods may offer interpretability, they are often constrained by oversimplified models that might sub optimally reflect real-world imaging systems. LensPlus instead leverages the flexibility of deep learning to learn mappings from real image data, capturing nuances of both system optics and biological variability. This makes our approach more adaptable and scalable.

That said, our current dataset consists of fixed HeLa cells, and thus does not capture the full range of biological diversity or imaging conditions. We present this study as a proof-of-principle demonstration of LensPlus, showing its potential in enhancing SBP using minimal hardware and real-world data. With appropriate retraining, the same framework can be investigated in future studies on diverse sample types, imaging modalities, and even live-cell imaging. For each new modality, the models need to be retrained. For a different sample within the same modality, if widely different in structure than the trained data, a transfer learning or fine tuning of the learned model on new data will be the most optimal solution. For live cell applications, we suggest training on fixed samples to avoid dynamics-induced pixel-level differences in the training pair.

Finally, while we report strong quantitative gains across multiple metrics (PSNR, MSSSIM, PCC, GraS-PSNR, GraS-MSSSIM, power spectral bandwidths, and FRC), we are cautious not to over-rely on single-number evaluations. Our results show that conventional pixel-wise metrics for our application degrade as image content increases--a limitation we address by introducing content-aware metrics and spectral domain analysis. The fact that our model shows consistent improvement in both spatial and spectral domains, while avoiding high-frequency hallucinations, reinforces the robustness and physical plausibility of the generated outputs.

In summary, LensPlus provides a practical, modular, and extensible solution for increasing resolution and SBP in optical microscopy without hardware modification. Its performance and grounding in realistic imaging constraints make it a compelling framework for next-generation computational microscopy.

## 5. Conclusion

In this study, we presented LensPlus, a deep learning–based framework for computationally enhancing the space-bandwidth product (SBP) of optical microscopy systems. Unlike traditional super-resolution approaches that rely on either hardware modification or strong physical priors, LensPlus offers a purely data-driven alternative that is flexible, practical, and adaptable to real-world imaging constraints. Using SLIM as the testbed, we demonstrated that LensPlus effectively enhances image resolution without hallucination while preserving structural fidelity, even when trained on paired data acquired with different numerical aperture objectives.

A key insight from this work is the recognition that low-NA and high-NA images represent fundamentally different optical sections of the sample due to differences in depth of field and sectioning. Despite this, LensPlus can learn a meaningful mapping across these imaging regimes, producing results that closely match the spatial and spectral characteristics of high-NA acquisitions without hallucinations. Our results confirm that a carefully trained deep network can overcome the limitations of naive pixel-wise mappings and deliver resolution gains grounded in physical plausibility.

Furthermore, we introduced content-aware evaluation metrics and spectral domain analyses that better capture the performance of resolution enhancement models in biological imaging. These metrics not only align better with visual perception but also expose the limitations of conventional metrics in content-varying datasets.

While our proof-of-concept implementation focused on SLIM images of fixed HeLa cells, the LensPlus framework is modality-agnostic and extensible to other imaging platforms such as fluorescence microscopy. With sufficient training data, it can be adapted to various sample types and imaging settings.

In conclusion, LensPlus represents a step forward in computational microscopy, offering a practical, modular, and adaptable (after retraining) solution to the longstanding trade-off between resolution and field of view. Its architecture and training strategy open new directions for data-driven enhancement in modern imaging, without increasing hardware complexity or cost, positioning it as a next-generation tool for biological and clinical imaging.

**Acknowledgement**

The authors acknowledge the support of funding agencies in conducting this research. We also acknowledge Dr. Young Jae Lee for preparing biological samples used in this study. Finally, we dedicate this work to the Late Dr. Gabriel Popescu.

**Code and data availability**

All the relevant data is included in the manuscript. Trained models, sample imaging data and the inference code are updated on GitHub (https://github.com/NehaRG-QPI/LensPlus). Full imaging data is not shared due to the size constraints but can be obtained upon reasonable request to the corresponding author.

**Funding**

M.A.A. was supported in part by NIH Award P41EB031772 (sub-project 6366).

**Disclosures**

The authors declare no conflict of interest.